\PassOptionsToPackage{table}{xcolor}
\documentclass[sigconf]{acmart}
\pdfoutput=1

\usepackage{booktabs} % For formal tables

\usepackage{epsfig,graphicx,subfigure}
\usepackage{epstopdf}

\usepackage{amsmath,amscd,amsbsy,amssymb,latexsym,mathrsfs,url}
\usepackage{enumitem,balance,mathtools}
\usepackage{wrapfig,subfigure}
\usepackage{mathrsfs, euscript}
\usepackage{multicol}
\usepackage{float}
\usepackage{enumitem}

\setcopyright{rightsretained}

\acmConference[Submitted to a confenrence]{ a Confenrence }{2017}{Canada}
\acmYear{2017}
\copyrightyear{2017}

\begin{document}
\title{Hidden Community Detection in Social Networks}

\author{Kun He}
\authornote{This work was done while the first author was a visiting professor at Cornell University.}
%\orcid{1234-5678-9012}
\affiliation{%
  \institution{Huazhong University of Science and Technology}
  \city{Wuhan}
  \country{China}
}
\email{brooklet60@hust.edu.cn}

\author{Yingru Li}
\authornote{Corresponding author.}
\affiliation{%
  \institution{Huazhong University of Science and Technology}
  \city{Wuhan}
  \country{China}
}
\email{szrlee@hust.edu.cn}

\author{Sucheta Soundarajan}
\affiliation{
  \institution{Syracuse University}
  \city{Syracuse}
  \state{NY}
  \country{USA}
}
\email{susounda@syr.edu}

\author{John E. Hopcroft}
\affiliation{
  \institution{Cornell University}
  \city{Ithaca}
  \state{NY}
  \country{USA}
  }
\email{jeh@cs.cornell.edu}

\begin{abstract}
We introduce a new paradigm that is important for community detection in the realm of network analysis.
Networks contain a set of strong, dominant communities, which interfere with the detection of weak, natural community structure.
When most of the members of the weak communities also belong to stronger communities, they are extremely hard to be uncovered.
We call the weak communities the \emph{hidden community structure}.

We present a novel approach called \emph{HICODE} (HIdden COmmunity DEtection) that identifies the hidden community structure as well as the dominant community structure. 
By weakening the strength of the dominant structure, one can uncover the hidden structure beneath. Likewise, by reducing the strength of the hidden structure, one can more accurately identify the dominant structure.
In this way, \emph{HICODE} tackles both tasks simultaneously.

Extensive experiments on real-world networks demonstrate that \emph{HICODE} outperforms several state-of-the-art community detection methods in uncovering both the dominant and the hidden structure. For example in the Facebook university social networks, we find multiple non-redundant sets of communities that are strongly associated with
residential hall, year of registration or career position of the faculties or students,
while the state-of-the-art algorithms mainly locate the dominant ground truth category.
Due to the difficulty of labeling all ground truth communities in real-world datasets, \emph{HICODE} provides a promising approach to pinpoint the existing latent communities and uncover communities for which there is no ground truth. Finding this unknown structure is an extremely important community detection problem.
\end{abstract}




\maketitle

\section{Introduction}
\label{sec:introduction}
Over the past decades, community detection has emerged as an essential task in the realm of network analysis,  
%Intuitively, a community is regarded as a set of nodes that is internally well-connected but comparatively less-connected to the rest of the network.
and provides insight into the underlying structure and potential functions of the networks~\cite{Girvan2002communities,Newman2003}.
%Diverse techniques are considered, including random walk diffusion~\cite{Rosvall2008Infomap}, heat kernel diffusion~\cite{HeatKernel2014}; modularity optimization~\cite{newman2004Mod,Newman2006Mod,blondel2008louvain}; %conductance optimization~\cite{HeICDM2015,AndersenWWW2006};
%seed set expansion~\cite{Whang2013seed, AndersenWWW2006}.% or local search~\cite{CuiSIGMOD2014,WuLocal2015}.
Early work focused primarily on identifying disjoint communities that partition the set of nodes within a network~\cite{Rosvall2008Infomap,blondel2008louvain,pons2006walktrap}. More recently, researchers have observed the multiplicity of interwoven memberships of communities
and have developed algorithms for finding overlapping communities~\cite{Ahn2010LinkCommunities,lancichinetti2011OSLOM,Yang2012ICDMb,coscia2012demon}.
Some partitioning techniques are also extended to tackle the overlapping case~\cite{Zhang2007Mod,He2015ICDM,Gregory2009Overlap}. %Li2015WWW
Within these two categories, one can further build a hierarchical dendrogram based on the granularity of the detected communities.
%Due to the topological nature, communities are naturally overlapped and tangled with each other.

Although much progress has been made, there is a type of community structure, which we call the \textbf{hidden community structure}, that has attracted little attention in the literature.
Real-world networks contain sparse community structure, such as secret organizations or temporary groups, which is considerably weaker than the dense community structure like families, colleagues or close friends, as evaluated by popular community scoring metrics.
If most of the members in the less modular community also belong to other denser communities, the community is usually overlooked.

\setlength{\belowdisplayskip}{-2pt} \setlength{\belowdisplayshortskip}{-2pt}
\setlength{\abovedisplayskip}{-2pt} \setlength{\abovedisplayshortskip}{-2pt}
\begin{figure*}[!htb]
	\subfigure[Dominant communities]{
		\label{fig:ExampleL1} %% label for first subfigure
		\begin{minipage}[b]{0.48\textwidth}
			\centering
			\includegraphics[height=1.8in]{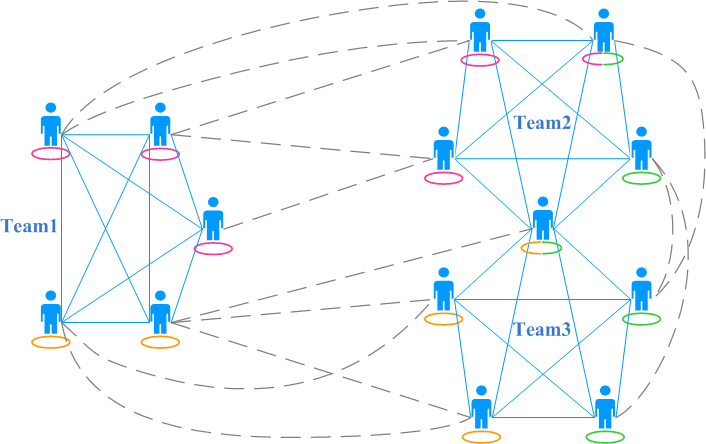}
	\end{minipage}}
	\subfigure[Hidden communities]{
		\label{fig:ExampleL2} %% label for second subfigure
		\begin{minipage}[b]{0.48\textwidth}
			\centering
			\includegraphics[height=1.8in]{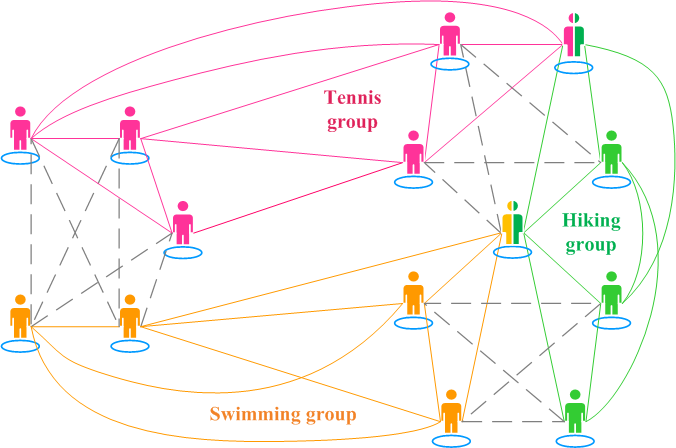}
	\end{minipage}}\vspace{-1em}
	\caption{The illustration of the dominant communities and the hidden communities in a social network. (a) The three cliques correspond to communities of students working closely as teams in projects. (b) The three groups of different colors correspond to sports communities with sparser connections.}
	\label{fig:HiddenExample}
\end{figure*}

For instance, in a social network, individuals may belong to multiple strong social communities, corresponding to groups such as families, colleagues and friends.  Though overlapping, the connections inside these communities are strong and numerous enough that existing overlapping community detection algorithms can perceive and uncover these latent but dominant modular structures.
However, in addition to these strong communities, individuals may also belong to weaker communities, such as a group of medical patients that see each other at the doctor's office and communicate infrequently, or high school alumni whom have infrequent contact.
As illustrated in Figure \ref{fig:HiddenExample} for a small network, the hidden community structure is sparser and as it is tangled with the structure of the dominant communities, it is also harder for detection.

For applications in a large variety of scientific disciplines, the weak, hidden structure is of great interest and deserves to be explored.
For example, in Protein-Protein Interaction (PPI) networks~\cite{park2011Isobase}, biologists wish to identify gene groups serving similar functions.
However, the current annotation is far from complete~\cite{HINT2012}, and the dominant, clearest groupings are more likely to be annotated. %park2011Isobase
In such a scenario, a way to help the biologists identify the hidden, less obvious groups is of great value.
As another example, consider a government that wishes to identify communities of a criminal or terrorist organization in a social network.
This community structure may be expressed within the network, but is likely to be much weaker than communities corresponding to family or location.
Identifying this hidden structure in the presence of the stronger community structure is even more important but faces a major challenge.

This paper aims to provide insight into the hidden structure. %develop a comprehensive understanding
We will address the following questions:

\begin{itemize}[noitemsep,nolistsep]
	\item Do real-world networks contain hidden community structure?
	\item How can we formally characterize ``hidden" structure? % What do me mean exactly by "hidden"?
	\item How can one find hidden structure in the presence of stronger, dominant structure?
	%\item Does the hidden structure interfere with the accurate detection of the dominant structure?%, and vise versa?
\end{itemize}

To address the above issues, we define the \textbf{hiddenness value} of a community as the portion of nodes in stronger communities,
and present a meta-approach called Hidden Community Detection (\emph{HICODE}) to identify the dominant structure as well as the hidden structure in networks.
\emph{HICODE} begins by first applying an existing algorithm as a base algorithm to a network, and then weakening the structure of the detected communities in the network. In this way, the weaker, hidden community structure becomes visible.  This step is repeated iteratively until no further significant structure is detected.  Next, \emph{HICODE} weakens the structure of the hidden communities, and thus obtains a more accurate version of the dominant community structure. %clearer picture of the dominant community structure. %original, dominant set.

Through experiments on real social networks, we demonstrate the existence of multiple sets of non-redundant communities.
The weaker set of communities, though possessing significant community structure, is rarely detected by state-of-the-art community detection algorithms.
On real-world networks containing a hidden ground-truth community structure,
\emph{HICODE} uncovers this structure much preciser than the baseline algorithms.

Hidden community structure can be regarded as a special type of overlapping communities. However, existing overlapping community detection methods mainly focus on communities in which a considerably portion of the members are not ``hidden", that is, they could also belong to other weaker communities but this community is clearly the strongest for these members.
We believe the insights we obtained on hidden community structure will provide valuable guidance for future investigations.
The main contributions of this paper include:
 
\begin{itemize}[noitemsep,nolistsep]
	\item \textbf{Conception on Hidden Community.} We introduce the concept of hidden community structure that exists widely in social networks, and we formally define the hiddenness value of a community.
	\item \textbf{Methods on Hidden Community Detection.} We present \emph{HICODE} for identifying both the dominant and the hidden structure. We implement \emph{HICODE} with several community detection algorithms as the base algorithm, and provide several structure weakening methods: RemoveEdge, ReduceEdge and ReduceWeight.
	\item \textbf{Validation on real world datasets.} Through experiments on a variety of real-world networks, we demonstrate that 
	the higher the hiddenness value a community is, the harder for an algorithm to locate such community; \emph{HICODE} outperforms several state-of-the-art community detection methods on uncovering the hidden communities.
	\item \textbf{Scalability.} As a meta-approach, \emph{HICODE} is scalable by applying any conventional community detection algorithm as the base algorithm. One can use \emph{HICODE} as a tool to find overlapping communities by using any disjoint community algorithm as the base. In addition, by applying overlapping community detection algorithms as the base, \emph{HICODE} can improve the accuracy on the set of communities the base algorithm found, at the same time it can uncover more weak communities hidden underneath the dominant set.
\end{itemize}

%\item We demonstrate that the presence of strong, dominant community structure interferes with the detection of the weaker, but still meaningful community structure, at the same time the hidden community structure can also hinder the accurate detection of the dominant structure.
% We provide justifications for \emph{HICODE} and explain why, under certain conditions, the weakening of one layer does not affect the quality of communities detected from another layer.
% and show that real-world networks from various domains contain multiple sets of non-redundant community structure that state-of-the-art community detection algorithms rarely uncover.
%\item  We show that through an iterative manner, \emph{HICODE} gradually removes the interplay among the multiple layers of communities, separates and finds high-modularity sets of communities in real networks across various domains, and uncovers the hidden communities. %outperforms popular community detection algorithms at the task of detecting ground truth communities, especially on the hidden communities.
%\item In contrast with related work that , \emph{HICODE} provides a much improved method of weakening the current detected community structure to find other weaker, hidden communities. More importantly, \emph{HICODE} uses a refinement process to greatly improve the detection quality of both the dominant communities and the hidden communities. %Add our progress as compared with Young's work, and the multi-view clustering methods.

\section{Preliminaries}
\label{sec:preliminaries}
 Let graph $G = (V, E)$ represent a network with $n$ nodes and $e$ edges.
 Let $A$ be the adjacency matrix of $G$, the $ij$-entry $A_{ij} \in \{0,1\}$ indicates whether there is an edge connecting nodes $i, ~j$.\footnote{For simplicity, we discuss unweighted graph, but all discussions in this paper are easily extended to weighted graph by letting $A_{ij} \in [0,1]$ to indicate the weight of each edge. In the weighted graph, the number of edges then corresponds to the sum of the edge weights.}
 Let $\mathcal{C} = \{C_1, C_2, ..., C_K\}$ be all the overlapping communities, where $C_k = (V_k, E_k)$ is a comparatively dense subgraph of $G$.
 The cardinality of a community is defined as the number of nodes in this community, i.e. $|C_k| = |V_k|$.
 We first introduce some necessary metrics, then provide formal definitions on the hidden structure.

 \subsection{Metrics}
 \label{subsec:Metrics}
 
 \textbf{Modularity.}
 To measure the strength of a set of communities that partition the network, we adopt the popular \emph{modularity} metric. Modularity is defined by Newman as the ratio of the number of intra-community edges to the expected number of edges in the same set of communities if the edges had been distributed randomly while preserving degree distribution~\cite{newman2004Mod}.
 
 The modularity score $Q$ of a partition is calculated by:
 \setlength{\belowdisplayskip}{0pt} \setlength{\belowdisplayshortskip}{0pt}
 \setlength{\abovedisplayskip}{0pt} \setlength{\abovedisplayshortskip}{0pt}
 \begin{equation}\label{EqMod}
 Q = \sum_{k=1}^K Q_k = \sum_{k=1}^K\left[\frac{e_{kk}}{e}-\left( \frac{d_k}{2e} \right)^2\right],
 \end{equation}
 where $K$ is the number of communities in the partitioning, $e$ is the number of edges in the graph, $e_{kk}$ is the number of edges within community $C_k$, and
 $d_{k}$ is the total degree of the nodes in community $C_k$.
 $Q$ lies in the range $[-0.5,1)$, with larger values indicating a stronger community set.
 
 \begin{definition}
 	\textbf{Modularity of a Community}. We call $Q_k$ the sum modularity contribution of $C_k$, and the modularity of a single community is defined as the sum modularity contribution of that community divided by the number of nodes in the community, i.e. $Q_k/|C_k|$.
 \end{definition}
 
 Zhang et al. extend Newman's definition to a set of overlapping communities by considering the belonging coefficient $w_{ik}$ for node $i$ to community $C_k$~\cite{Zhang2007Mod}. Briefly, $w_{ik} = 1/m$ if community $C_k$ is one of the $m$ communities containing node $i$. Then in Eq. (\ref{EqMod}), $e_{kk}$ is weighted by  
 \begin{equation}\label{Eq22}
 e_{kk} = \frac{1}{2} \sum_{i,j \in C_k} \frac{w_{ik}+w_{jk}}{2} A_{ij},
 \end{equation}
 where $A_{ij}$ is the $ij$-entry of the adjacency matrix. We know $d_k = 2e_{kk}+e_{k\_out}$, then $e_{k\_out}$ is weighted by
 %$e_{k\_out} = \frac{1}{2} \sum_{i \in C_k, j \notin C_k} \frac{w_{ik}+(1-w_{jk})}{2} A_{ij}.$
 %\setlength{\belowdisplayskip}{0pt} \setlength{\belowdisplayshortskip}{0pt}
 %\setlength{\abovedisplayskip}{0pt} \setlength{\abovedisplayshortskip}{0pt}
 \begin{equation}\label{Eq23}
 e_{k\_out} = \frac{1}{2} \sum_{i \in C_k, j \notin C_k} \frac{w_{ik}+(1-w_{jk})}{2} A_{ij}.
 \end{equation}
 
 The extended definition degenerates exactly to Eq. (\ref{EqMod}) for disjoint communities.
 
 %\textbf{Conductance.}
 %The conductance of a single community $C$ is defined as the fraction of total edge volume leaving the community\cite{Ncut2000}.
 %\setlength{\belowdisplayskip}{0pt} \setlength{\belowdisplayshortskip}{0pt}
 %\setlength{\abovedisplayskip}{0pt} \setlength{\abovedisplayshortskip}{0pt}
 %\begin{equation}\label{EqCond}
 %\phi_c =\frac{e_{c\_out}}{\min(d_c,d_{\bar{c}} )},
 %\end{equation}
 %where $e_{c\_out}$ denotes the cut size, and $d_c$, $d_{\bar{c}}$ the sum of node degree in $C$ and $\bar{C}$.
 
 \textbf{Normalized Mutual Information (NMI).}
 We use \emph{normalized mutual information (NMI)} to capture the similarity of two partitions $X$ and $Y$~\cite{Danon2005NMI}.
 \setlength{\belowdisplayskip}{0pt} \setlength{\belowdisplayshortskip}{0pt}
 \setlength{\abovedisplayskip}{0pt} \setlength{\abovedisplayshortskip}{0pt}
 \begin{equation}
 NMI(X,Y)=\frac{2I(X,Y)}{H(X)+H(Y)},
 \end{equation}
 where $H(X)$ is the Shannon entropy of partition $X$, and $I(X,Y)$ is the \emph{mutual information} that captures the similarity between two partitions $X$ and $Y$.
 For a set of communities that partition a network,
 $p(x) = |x|$ %\footnote{|x| denotes the number of vertices in community $x$.}%
 , and $p(x,y) = |x \cap y|$.   
 The NMI score lies in the range [0,1], where 1 represents a perfect matching and 0 indicates total independence.
 For overlapping communities, we use the extended NMI~\cite{overlapNMI}. See \cite{overlapNMI} for details.
 \begin{align}
 I(X,Y)	& = \sum_{x\in X}\sum_{y\in Y}p(x,y)\log\frac{p(x,y)}{p(x)p(y)} \\
 H(X) 	& = - \sum_{x\in X} p(x)\log p(x)
 \end{align}
% \begin{equation}
% I(X,Y) = \sum_{x\in X}\sum_{y\in Y}p(x,y)\log\frac{p(x,y)}{p(x)p(y)} 
% \end{equation}
% \begin{equation}
% H(X)  = - \sum_{x\in X} p(x)\log p(x)
% \end{equation}

 \subsection{Definitions}
 \label{subsec:Definition}
 Assume we have some metric function $\mathcal{F}$:($G, C_k$)$\mapsto \mathbb{R}$ (\emph{e.g., modularity~\cite{newman2004Mod} or conductance~\cite{Ncut2000}}) that assigns a quality score to a community.
 For simplicity, let $\mathcal{F}_k$ denote the quality of $C_k$ in $G$.  Here, we assume that higher scores indicate stronger communities, but the definition below can be trivially modified for the case when lower scores indicate higher community quality.
 %We define the hiddenness value of a community:
 %\vspace{-0.5em}
 \begin{definition}
 	\textbf{Hiddenness Value of a Community}. The hiddenness value $H(C_{k})$ of community $C_k$ is the fraction of nodes of $C_k$ belonging to various communities with a higher $\mathcal{F}$ score.
 \end{definition}
 %\vspace{-0.5em}
 Let $\mathbb{S}_k$ be the set of all strong communities for community $C_k$.
 \setlength{\belowdisplayskip}{0pt} \setlength{\belowdisplayshortskip}{0pt}
 \setlength{\abovedisplayskip}{0pt} \setlength{\abovedisplayshortskip}{0pt}
 \begin{equation}
 \mathbb{S}_k = \{ C_i | \mathcal{F}_i > \mathcal{F}_k, C_i \in \mathcal{C} \}
 \end{equation}
 
 The hiddenness value of $C_k$ can be calculated as:
 \setlength{\belowdisplayskip}{0pt} \setlength{\belowdisplayshortskip}{0pt}
 \setlength{\abovedisplayskip}{0pt} \setlength{\abovedisplayshortskip}{0pt}
 \begin{equation}
 %H(C_k) = \frac{ \bigg| \displaystyle\bigcup_{C_i \subseteq \mathbb{S}_k} C_i \cap C_k  \bigg| }{|C_k|},
 H(C_k) = \frac{1}{|C_k|} \cdot \bigg| \displaystyle\bigcup_{C_i \in \mathbb{S}_k} C_i \cap C_k  \bigg|
 \end{equation}

 $H(C_k) \in [0,1]$. Intuitively, the higher a community's hiddenness value is, the more difficult for the community to be uncovered.
 
 The goal of the hidden community detection problem is to locate overlapping communities in the network such that communities having high hiddenness values can be found. Note that there is no single threshold for a `high' hiddenness value, these values depend on the network under study, the metric being used, and the set of communities.
 For two communities $C_i, C_j \in \mathcal{C}$, if $H(C_i) > H(C_j)$, then $C_i$ is comparatively hidden as compared with $C_j$, and $C_j$ is comparatively dominant as compared with $C_i$.

 For convenience, we separate the communities into layers.
 
 \begin{definition}                                                                                                                        
 	\textbf{Layer}. A layer is a set of communities that partitions or covers the nodes of the network.\footnote{Here we allow trivial communities of size less than 3.}
 \end{definition}
 
 Intuitively, in a social network, a partitioning layer may correspond to a grouping of individuals by their locations,
 and colleague circles could form an overlapping layer where some colleagues are in interdisciplinary areas.
 \begin{definition}
 	\textbf{Hiddenness Value of a Layer}. The hiddenness value $H(\mathcal{L}_i)$ of a layer $\mathcal{L}_i$ is the weighted average hiddenness values of the communities in this layer.
 	\begin{equation}
 	%H(\mathcal{L}_i) = \sum_{C_k \subseteq \mathcal{L}_i} \frac{ |C_k|\cdot H(C_k)} {  \bigg| \sum_{C_j \subseteq \mathcal{L}_i}  C_j  \bigg|   }
 	H(\mathcal{L}_i) =  \frac{ \sum_{C_k \in \mathcal{L}_i}|C_k|\cdot H(C_k)} {  \bigg| \sum_{C_k \in \mathcal{L}_i}  C_k  \bigg|   }.
 	\end{equation}
 \end{definition}
 $H(\mathcal{L}_i) \in [0,1]$.
 The dominant layer is the layer with lowest hiddenness value. It is usually the set of communities found by a standard community detection algorithm that optimizes metric $\mathcal{F}$. A layer is called hidden if it has a comparatively high hiddenness value. 
 Figure \ref{fig:HiddenExample} illustrates a small social network with two layers of communities, the dominant layer corresponds to project teams and the hidden layer corresponds to sports groups.
 
 \vspace{-0.5em}
 \subsection{Case Study}
 
 To illustrate the concept of hidden community structure, we choose modularity as the community metric, and show an example on the Caltech Facebook network dataset (\emph{described in detail in Section~\ref{sec:realDatasets}}).
 For simplicity, here we only consider two annotations for each individual: the residence hall (`Dorm') and year of matriculation (`Year').
 A ground truth community contains all individuals living in the same dormitory, or are in the same year of matriculation.
 `Dorm' communities partition the nodes and form a layer. Similarly, `Year' communities form another layer.
 
 %% Calculated by Yingru
 %For Rice:
 %Avg. fraction of dorm communities hidden by a year community = 0.0215 (using mod), 0.2980 (using conductance)
 %Avg. fraction of year communities hidden by a dorm community = 0.9140 (using mod), 0.6397 (using conductance)
 %
 %For Caltech:
 %Dorm hidden by year = 0.0788 (using mod), 0.3818 (using conductance)
 %Year hidden by dorm = 0.7861 (using mod), 0.5057 (using conductance)
 
 On average, 79\% of the nodes in each `Year' community belong to a stronger `Dorm' community, i.e., the `Year' layer has a hiddenness value of 0.79. In contrast, the `Dorm' layer only has a hiddenness value of 0.08.
 %If we measure strength using conductance, then the hiddenness value of `Dorm' and `Year' are 0.38 and 0.50 respectively, and `Year' is still the hidden layer.
 The `Year' layer have substantially higher hiddenness values than the `Dorm' layer. Note that `Year' is still the hidden layer as compared with `Dorm' when we use the conductance metric.
 
 We thus conjecture that standard community detection algorithms have difficulty detecting the `Year' layer.  As we will see in the experiments,
 the state-of-the-art community detection algorithms that we consider can effectively find the `Dorm' layer but rarely detect the `Year' layer,
 while \emph{HICODE} can effectively uncover both layers.

\vspace{-0.5em}
\section{Hidden Community Detection}
\label{sec:Method}

\subsection{Algorithm Overview}

In this section, we propose a meta-approach called \emph{HICODE} to find layers of community structure in a network. \emph{HICODE} uses an existing community detection algorithm as the `base' method, and iteratively weakens the structure of detected layers to reveal hidden structure. \emph{HICODE} then applies a refinement procedure to increase the quality of the layers.

\emph{HICODE} contains two stages: \textbf{Identification} and \textbf{Refinement}.\\
\vspace{0.5em}
\textbf{Stage 1. Identification:}

The Identification stage determines the initial layers of communities as follows:

\begin{enumerate}[noitemsep,nolistsep]
	\item Identify a layer of communities via the base method;
	\item Weaken the structure of the detected layer;
	\item Repeat until the appropriate number of layers are found.  
\end{enumerate}

In this stage, we iteratively identify a set of initial layers by weakening the structure of the previous, stronger layers.

Section~\ref{sec:reduceMethods} presents methods for weakening the detected community structure in order to reveal the hidden structure beneath.

A crucial aspect of the Identification stage is to automatically determine the number of layers $n_L$ in a network.
This is accomplished by increasing $n_L$ until a stopping condition is met. We describe this process in Section~\ref{sec:num_layers}.

\vspace{0.5em}
\textbf{Stage 2. Refinement:}

It is reasonable that stronger community structure can obscure weaker community structure, but critically, we observe that weaker structure can also hinder the accurate or complete detection of the stronger structures.  After the \mbox{Identification} stage, one has only a rough approximation of the various community layers, and the purpose of the Refinement stage is to further improve the quality of these detected layers.

Refinement is an iterative process.  In each iteration, we consider each layer, and improve the current layer as follows:

\begin{enumerate}[noitemsep,nolistsep]
	\item Weaken the structures of \textit{all other} layers from the original network to obtain a \textit{reduced network};
	\item Apply the base algorithm to the resulting network.
\end{enumerate}

In contrast to the Identification stage, where only the layers found so far (i.e., the stronger layers) are reduced, during the Refinement stage, we weaken the effects of both the stronger and weaker layers.
This is necessary because the weaker layers can impair detection of the layer currently under consideration,
even though they have a smaller impact on the network structure than the stronger layers.  Through this process, a more accurate version of the current layer is produced.

We use NMI~\cite{Danon2005NMI,overlapNMI} to capture the similarity of two partitions or two sets of overlapping communities.
For the synthetic data that we consider, the Refinement stage quickly converges within 30 iterations, meaning that the NMI of the corresponding layers at adjacent iterations is almost 1.  For real-world datasets, 100 iterations is generally enough.

Although the trend over these iterations is to see higher quality community layers (i.e., higher modularity partitions or coverings), there are fluctuations in this trend.  In each Refinement iteration, we calculate the modularity of the detected layers, and the final output corresponds to the iteration with the highest average modularity score.

\vspace{-1em}
\subsection{Reducing Methods}
\label{sec:reduceMethods}

We present the following methods to reduce a single layer of community structure: RemoveEdge, ReduceEdge, and ReduceWeight.
RemoveEdge works reasonably well when there are only a small number of layers, and communities in different layers have comparatively small overlaps.
ReduceEdge generally performs better, but it typically does not perform as well as ReduceWeight; however, one major advantage that ReduceEdge has over ReduceWeight is that it does not require a base algorithm that supports weighted networks.
As we will see in experiments, ReduceWeight is the best-performing method of the three.

For all cases, if one has a layer of overlapping communities, then in order to avoid duplicate weakening on the overlapping portions, we sort the communities according to their sizes,
weaken larger communities first and do not weaken the overlapping portion again for subsequent communities in the same layer.

\subsubsection{RemoveEdge.}
RemoveEdge weakens a detected community by removing all intra-community edges.
This method is inspired by algorithms that find communities by removing edges, such as the classic Girvan-Newman algorithm, which, removes edges with high betweenness~\cite{Girvan2002communities}.  
RemoveEdge works reasonably well when there are few layers and communities in different layers have comparatively small overlaps. 
\subsubsection{ReduceEdge.}
ReduceEdge approximates each layer as a single stochastic blockmodel, where other edges are regarded as background noise.  This method randomly removes some edges within each community block so that the edge probability in the block matches the background edge probability of this block.

\begin{figure}[htbp]
	\vspace{-0.5em}
	\centering
	\includegraphics[width=1in]{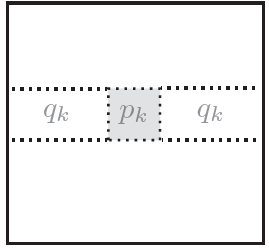}
	\vspace{-3mm}
	\caption{$p_k$ and $q_k$ of a community block $C_k$.}
	\label{reducemethod}
\end{figure}

For each community $C_k$ in a single layer that we want to reduce, we
calculate the observed edge probability $p_k$ in $C_k$, and the background block probability $q_k$, as illustrated in the reordered adjacency matrix in Figure \ref{reducemethod}.
Let $n_k$ and $e_{kk}$ be the number of nodes and edges inside $C_k$,
and $d_k$ the sum of degrees of nodes in $C_k$.
$p_k$ equals the actual number of edges in $C_k$ divided by the maximum possible number of edges:
\setlength{\belowdisplayskip}{1pt} \setlength{\belowdisplayshortskip}{1pt}
\setlength{\abovedisplayskip}{1pt} \setlength{\abovedisplayshortskip}{1pt}
\begin{equation}
p_k = \frac{e_{kk}}{0.5n_k(n_k-1)},
\end{equation}
and $q_k$ is the average outgoing edge density of block $C_k$:
\setlength{\belowdisplayskip}{1pt} \setlength{\belowdisplayshortskip}{1pt}
\setlength{\abovedisplayskip}{1pt} \setlength{\abovedisplayshortskip}{1pt}
\begin{equation}
q_k = \frac{d_k - 2e_{kk}}{n_k(n - n_k)}.
\end{equation}

\vspace{-0.5em}
If we treat the observed edge probability $p_k$ in community $C_k$ as the superposition of the underlying edge probability $p'_k$ of $C_k$ and the background block probability $q_k$, then $p_k = 1 - (1-p'_k)(1-q_k)$.
For an edge in $C_k$, the probability that it is generated by the background noise is
\setlength{\belowdisplayskip}{0pt} \setlength{\belowdisplayshortskip}{0pt}
\setlength{\abovedisplayskip}{0pt} \setlength{\abovedisplayshortskip}{0pt}
\begin{equation}
q'_k = 1 - p'_k = \frac{1-p_k}{1-q_k} = \frac{q_k}{p_k}.
\label{Eq:qk}
\end{equation}

ReduceEdge removes each edge within community $C_k$ with probability $1 - q'_k$ (i.e., it keeps each internal edge
with probability $q'_k$).  In this way, edges are randomly removed from $C_k$ such that the edge probability within $C_k$ matches $q_k$.\footnote{Note that ReduceEdge is also suitable for weighted graph where $e_{kk}$ corresponds to the sum of edge weights inside $C_k$, and $d_k$ corresponds to the sum of degrees of nodes in $C_k$, weighted by the edge weight.}

\subsubsection{ReduceWeight.}
This method reduces the weight of each edge within community $C_k$ by a factor of $q'_k$, defined in Eq. (\ref{Eq:qk}). Like with ReduceEdge, we wish to set the weighted probability within $C_k$ equal to the average weighted background block probability $q_k$.

Note that to use ReduceWeight, one's base algorithm must support weighted networks.  However, the original network itself need not be weighted, as one can simply set the original weight of every edge to 1.
We let ReduceWeight degenerate to ReduceEdge if the base algorithm does not support weighted graphs.
Unlike ReduceEdge, ReduceWeight is deterministic.

\subsection{Selecting the Number of Layers}
\label{sec:num_layers}
A major challenge for \emph{HICODE} is determining $n_L$, the appropriate number of community layers.
On synthetic networks (\emph{details of the data model are described Section \ref{sec:blockmodels}}) where all communities are labeled, we observe that if $n_L$ is chosen correctly, then during the Refinement stage, the average modularity of the detected layers increases. If $n_L$ is either over or under-estimated, then this trend declines. Our rule for determining the number of layers $n_L$ is motivated by this observation: if one selects the appropriate number of layers, the output will generally be of a higher quality.

We begin by setting the number of layers $n_L = 2$, and increase the number of layers until a stopping condition is met.   
For each candidate number of layers $n_L$, \emph{HICODE}
first calculates the modularity of the weakest layer obtained at the Identification stage.
If the value is very low, then there are no more significant layers, so we set $n_L = n_L - 1$ and return;
otherwise, let $Q_t$ be the average modularity of all the detected layers at step $t$:

\begin{enumerate}[noitemsep,nolistsep]
	\item  Calculate $Q_0$ for $t = 0$, i.e. after identification, before any refinement is conducted;
	\item  Perform $T = 10$ tentative iterations of refinement, and calculate $Q_t$ for each $t \in \{1,...,T\}$;
	\item Calculate the average improvement ratio of modularity per iteration\footnote{Instead of only considering how much improvement we get at step $T$,
		we use the average modularity of the $T$ steps to balance the fluctuation on real-world networks.}: $R_T = \frac{\sum_{t=1}^T Q_t}{T \cdot Q_0}$.
\end{enumerate}

$R_T$ represents how much refinement improves the detected layers.
If $n_L$ is too high, then when we remove the structure of the extra layers,
we will be removing structure that actually belongs to some earlier layer.
This will result in a lower quality partition, such that the refinement stage actually \textit{lowers} the quality of the detected layers.
Thus, we choose $n_L$ corresponding to the peak $R_T$. % before descending.

\section{Experimental Setup}
\label{sec:ExperimentalSetup}

\vspace{-0.3em}
\subsection{Evaluation Metric}
\label{sec:eval_methods}

Besides $NMI$, we also define a Jaccard score-based metric to evaluate how well the detected layers resemble the ground truth communities.  
Given a set of detected communities $\mathcal{D}$ and a set of ground truth communities $\mathcal{G}$,
the Jaccard similarity-based precision, recall, and $F_1$ score are defined as follows:

Each detected community $D_i$ has its individual Jaccard Precision
$$P(D_i) = \mathop{\max}\limits_{G_j \in \mathcal{G}}\frac{|G_j\cap D_i|}{|G_j\cup D_i|}.$$
The \emph{Jaccard Precision $P$} of the set $\mathcal{D}$ is defined as the weighted average of $P(D_i)$ over all detected communities, weighted by the size of the communities.

Each ground truth community $G_j$ has its individual Jaccard Recall
$$R(G_j) = \mathop{\max}\limits_{D_i \in \mathcal{D}}\frac{|G_j\cap D_i|}{|G_j\cup D_i|}.$$
The \emph{Jaccard Recall $R$} of the set $\mathcal{G}$ is defined as the weighted average of $R(G_j)$ over all ground truth communities,
weighted by the size of these communities.

We use weighted $P$ and $R$ to give a higher priority to bigger communities. The
\emph{Jaccard $F_1$ Score} is defined as the harmonic mean of $P$ and $R$, i.e. 
$$F_1 = 2PR/(P+R).$$

\vspace{-0.3em}
\subsection{Synthetic Stochastic Blockmodel}
\label{sec:blockmodels}

We define a synthetic stochastic blockmodel containing multiple layers of planted communities.
Each layer in a network corresponds to a single stochastic blockmodel that partitions the nodes into roughly equally-sized sets.
In each layer, we first create the appropriate number of community IDs and randomly assign each node to a community, then we produce a $G(n, p)$ Erdos-Renyi random graph over each block.  We select suitable $p$-values for the different layers so that they are of different densities, but roughly equal strengths as measured by modularity.
By randomly assigning nodes to communities, we ensure that the communities in different layers are independent (i.e., a node's membership in different layers is unrelated).

We then generate a small synthetic network SynL2 that will be used to illustrate the details of our algorithm.
There are two layers of communities, containing 5 communities of roughly size 40, and 4 communities of roughly size 50.  The $p$ values that govern the number of intra-community links for these two layers are $0.12$ and $0.10$. Their modularity scores are $0.40$ and $0.39$, with hiddenness values of 0.43 and 0.56. We observe a very low $NMI$ of 0.05 between the two layers, and the $F_1$ score is also low at 0.02.

\vspace{-0.3em}
\subsection{Real-world Datasets}
\label{sec:realDatasets}
We do thorough testing on two groups of real-world networks, as summarized in Table~\ref{table:network_stats}.

(1) \textbf{Facebook networks}:  These networks are portions of the Facebook social network for different universities in the United States~\cite{facebook100}, which comes from a single-day snapshot in September 2005 when one needed a `.edu' email address to become a member of Facebook.
For each university network, the data have categorical attributes encompassing the gender, major, year of matriculation, high school, dormitory, status (faculty, student, staff, etc.) and residence (house, dormitory, fraternity, etc.) of the users. We choose eight representative university networks which express comparatively high modularity scores when grouping the nodes by at least one of the attributes: Caltech, Smith, Rice, Vassar, Wellesley, Bucknell, Carnegie and UIllinois.

(2) \textbf{SNAP networks}: We choose three networks with ground-truth communities collected by SNAP\footnote{http://snap.stanford.edu}~\cite{Jaewon2012ground}:
Youtube, Amazon and DBLP.
Youtube indicates friendships among users and the ground truth communities are user-defined groups.
Amazon is a product co-purchasing network where co-purchased products are connected by edges, and
the ground truth corresponds to product categories. 
DBLP is a co-authorship network where nodes represent authors, edges represent the co-authorship, and the ground truth communities correspond to conferences.
For each network, we extract a subnetwork (\emph{For convenience, we still use the same name}) containing all nodes in the top 5000 ground-truth. 
For further analysis, we partition the ground-truth communities into several community sets ($CommSet$) based on their hiddenness values.

\vspace{-0.3em}
\begin{table}[htp]
	\centering
	\small
	\scalebox{0.9}{
		\begin{tabular}{l l l  r r }
			\hline
			\bf{Source} &\bf{Domain} &\bf{Dataset} &\bf{$|V|$} & \bf{$|E|$} \\
			\hline
			\bf{Facebook}&\bf{Social}&Caltech	& 769	    	& 16,656 				\\
			&	         &Smith		& 2,970 		& 97,133      			\\
			&            &Rice		& 4,087 		& 184,828       		\\
			&	         &Vassar	& 3,068 		& 119,161      			\\
			&            &Wellesley	& 2,970 		& 94,899       			\\
			&	         &Bucknell 	& 3,826 	    & 158,864 				\\
			&            &Carnegie	& 6,637 		& 249,967       		\\
			&            &UIllinois	& 30,809		& 1,264,428       		\\
			\hline
			\bf{SNAP}	& \bf{Social}       &YouTube	& 31,150	& 202,130	    \\
			& \bf{Products}	    &Amazon		& 13,288	& 41,730		\\
			& \bf{Collaboration}&DBLP 		& 49,097	& 170,284		\\
			\hline
	\end{tabular}} \vspace{0.1em}
	\caption{The real-world network datasets.}
	\label{table:network_stats}
	\vspace{-2.5em}
\end{table}

More statistics for their ground truth communities are as shown in Table~\ref{table:all_network_layers} (left column): the modularity and average hiddenness value for each ground truth community layer, as well as the maximum $NMI$ ($NMI_{max}$), maximum and average $F_1$ scores ($F_{max}$ and $F_{avg}$) for pairwise layers.

For Facebook networks, each layer gives rise to a set of communities grouped by a common attribute
(i.e., nodes with a common annotation are in the same community), and we call each such set of communities an annotated layer
(e.g., the `Dorm' annotation gives rise to one annotated layer).
These annotated layers cover all nodes ($coverage=100\%$) in the corresponding networks.

For SNAP networks, we manually divide the ground-truth communities into two sets, namely $CommSet_A$ or $CommSet_B$, depending on whether the hiddenness value is less than 0.5 or not. 
The coverage ratio (\textit{ratio of nodes covered by the communities}) of $CommSet_A$ in Amazon, DBLP and Youtube are 100\%, 100\% and 80\%. So $CommSet_A$ forms a layer for Amazon and DBLP respectively, but it only covers 80\% of the nodes in Youtube. The coverage ratio of $CommSet_B$ on the three networks are 46\%, 4\% and 46\% respectively.
$CommSet_A$ and $CommSet_B$ contain very different communities for DBLP and Youtube. 
But on Amazon, $CommSet_A$ and $CommSet_B$ contain very similar communities since their pairwise $NMI$ and $F_1$ are very high, indicating that if an algorithm accurately detects most communities in $CommSet_A$,  then this algorithm also has a high detection accuracy on  $CommSet_B$.

\vspace{-0.8em}
\subsection{Base Algorithms and Baselines}

We compare \emph{HICODE} to four popular algorithms in two categories:

(1) \textbf{Overlapping detection methods}: OSLOM (OS)~\cite{lancichinetti2011OSLOM} and Link Communities (LC)~\cite{Ahn2010LinkCommunities}. OS uses a fitness function and joins together small clusters into statistically significant larger clusters. LC is a landmark algorithm that finds communities by performing hierarchical clustering on the links, which results in overlapping communities of nodes.

(2) \textbf{Disjoint detection methods}: Infomap (IM)~\cite{Rosvall2008Infomap} and Louvain method (Mod)~\cite{blondel2008louvain}. IM is based on the random walk technique and minimizes the expected length of a description of information. Mod is a fast method popular for greedy modularity optimization.

We also implement \emph{HICODE} using the four algorithms as the base, denoted as {\sc HC:Mod}, {\sc HC:IM}, {\sc HC:OS}, and {\sc HC:LC}.\footnote{Code and synthetic data: https://github.com/KunHe2015/HiCode/}
% In general, {\sc HC:Mod} performs consistently better than {\sc HC:IM}, and {\sc HC:OS} performs consistently better than {\sc HC:LC}. 

\section{Experimental Results}
\label{sec:ExperimentalResults}

We first evaluate different reducing methods of \emph{HICODE} and illustrate the necessity of the Refinement stage. 
Then we show the statistics of the multiple layers detected by \emph{HICODE} and compare the detection accuracy with state-of-the-art baselines on real-world networks.

%Experiments show that \emph{HICODE} finds multiple significant, non-redundant community layers. These community layers are of high quality when evaluated mathematically by the popular modularity metric.
%Many of the layers are strongly associated with annotated communities grouped by living residence, year of registration or career position.  The weaker layers are masked by the stronger layers and they are rarely uncovered by existing community detection algorithms.

\vspace{-0.5em}
\subsection{Analysis on HICODE}
\label{sec:experimentsHC}
\subsubsection{Comparison of the Reduction Methods}
\begin{figure*}[htbp]
	\vspace{-0.6em}
	\subfigure[Caltech]{
		\label{fig:reduceMethod_Caltech} %% label for second subfigure
		\begin{minipage}[b]{0.22\textwidth}
			\includegraphics[width=1.5in]{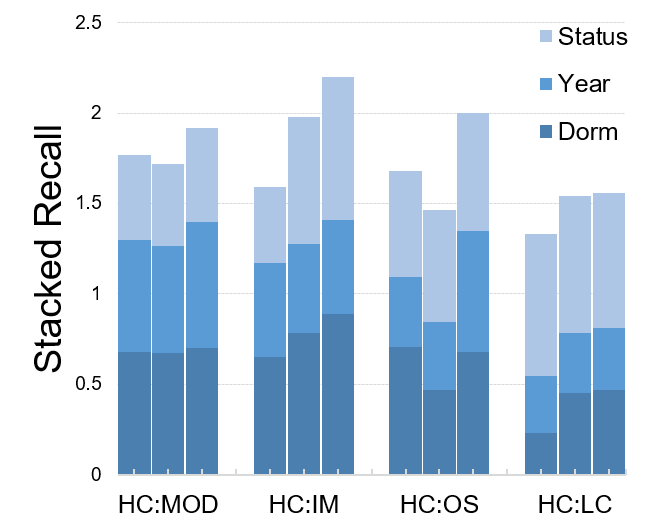}
	\end{minipage}}
	\subfigure[Smith]{
		\label{fig:reduceMethod_Smith} %% label for second subfigure
		\begin{minipage}[b]{0.22\textwidth}
			\includegraphics[width=1.5in]{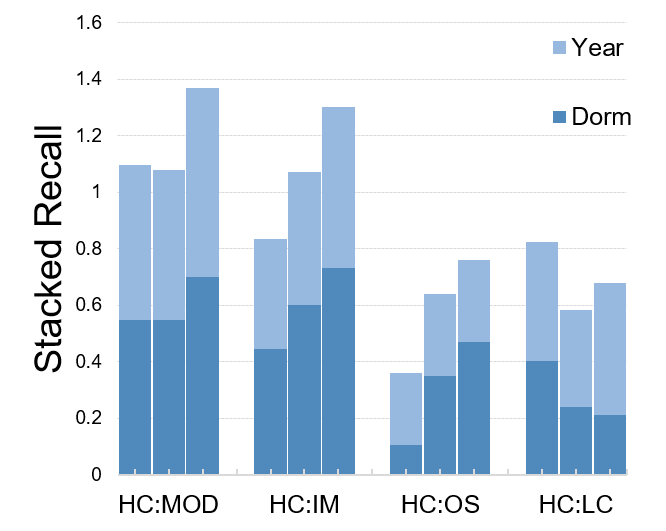}
	\end{minipage}}
	\subfigure[Rice]{
		\label{fig:reduceMethod_Rice} %% label for second subfigure
		\begin{minipage}[b]{0.22\textwidth}
			\includegraphics[width=1.5in]{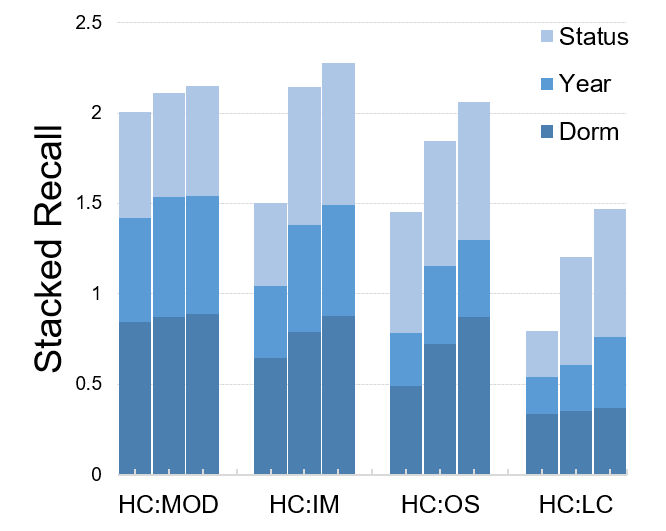}
	\end{minipage}}
	\subfigure[Vassar]{
		\label{fig:reduceMethod_Vassar} %% label for second subfigure
		\begin{minipage}[b]{0.22\textwidth}
			\includegraphics[width=1.5in]{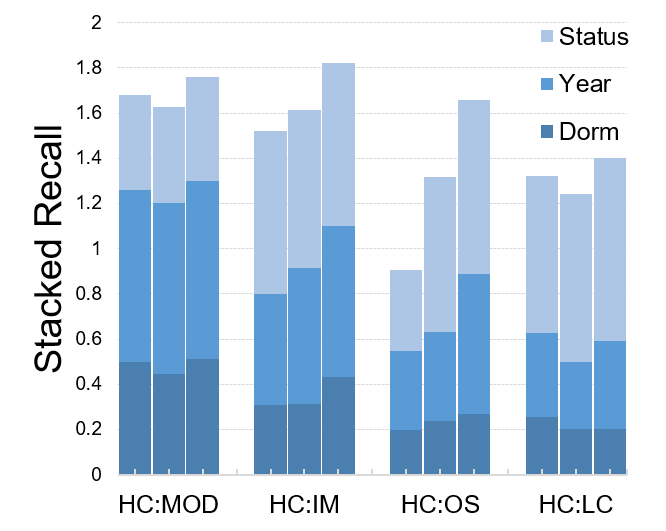}
	\end{minipage}}
	\vspace{-1.5em}
	\caption{Stacked Recall of all Layers for the reducing methods (left to right: RemoveEdge, ReduceEdge, ReduceWeight).}
	\vspace{-0.3cm}
	\label{fig:Reduce_Method} %% label for entire figure
\end{figure*}

We first evaluate different reducing methods on each of the versions of \emph{HICODE} (corresponding to different base algorithms).
Figure \ref{fig:Reduce_Method} shows the Recall for all layers of communities detected by \textbf{{\sc HC:Mod}}, \textbf{{\sc HC:IM}}, \textbf{{\sc HC:OS}} and \textbf{{\sc HC:LC}} on several examples of the Facebook networks: Caltech, Smith, Rice and Vassar.

\setlength{\belowdisplayskip}{0pt} \setlength{\belowdisplayshortskip}{0pt}
\setlength{\abovedisplayskip}{0pt} \setlength{\abovedisplayshortskip}{0pt}
\vspace{-0.1em}

In general, ReduceWeight reaches the highest detection accuracy, followed by ReduceEdge and RemoveEdge.
ReduceWeight and ReduceEdge provide a more accurate estimation on the background density,
while RemoveEdge removes all intra-community edges and impairs more structure of other layers, especially when communities in different layers overlap considerably.
ReduceWeight deterministically reduces the weight of intra-community edges and outperforms ReduceEdge, which randomly removes intra-community edges.
To save space, we will present results using ReduceWeight.

For the different versions of \emph{HICODE}, {\sc HC:Mod} and {\sc HC:IM} perform consistently better than {\sc HC:OS} and {\sc HC:LC}. We will use {\sc HC:Mod} to show the property of \emph{HICODE}, and then compare our results on the four \emph{HICODE} implementations with other algorithms.
%\vspace{-0.5em}
\subsubsection{Necessity of the Refinement Stage.}
To show the necessity of the Refinement stage,
which gradually separates the community structures mixed together and strengthens the structure of each layer,
we run {\sc HC:Mod} on the small network SynL2 and illustrate its results.\footnote{The other base algorithms produce similar results.}

\begin{figure}[htbp]%[t!]
%	\vspace{-1em}
	\subfigure[$t = 0$]{
		\begin{minipage}[b]{0.105\textwidth}
			\centering
			\includegraphics[width=0.725in]{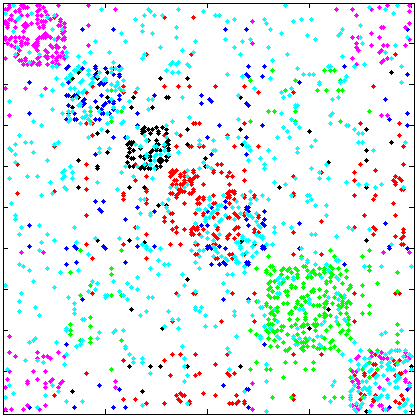}
	\end{minipage}}
	\subfigure[$t = 5$]{
		\begin{minipage}[b]{0.105\textwidth}
			\centering
			\includegraphics[width=0.725in]{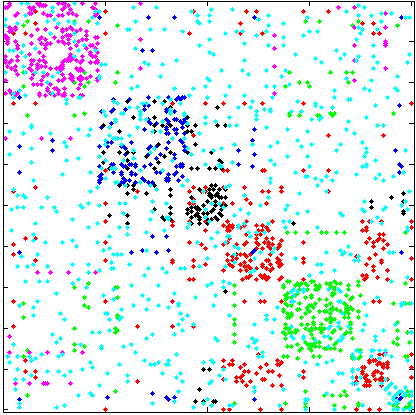}
	\end{minipage}}
	\subfigure[$t = 10$]{
		\begin{minipage}[b]{0.105\textwidth}
			\centering
			\includegraphics[width=0.725in]{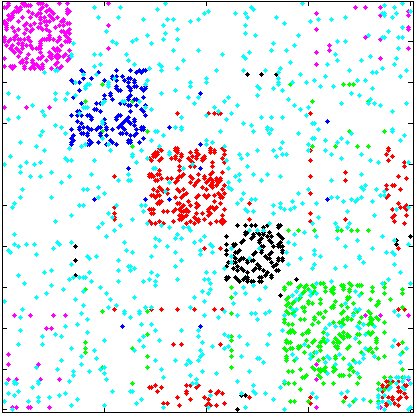}
	\end{minipage}}
	\subfigure[$t = 20$]{
		\begin{minipage}[b]{0.105\textwidth}
			\centering
			\includegraphics[width=0.725in]{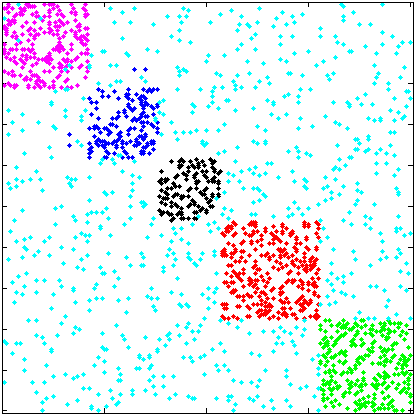}
	\end{minipage}} \vspace{-1.5em}
	\caption{Refinement of layer 1 on SynL2.}
	\label{fig:SynL2_200_layer1} %% label for entire figure
\end{figure}

\begin{figure}[htbp]%[t!]
	\vspace{-1em}
	\subfigure[$t = 0$]{
		\begin{minipage}[b]{0.105\textwidth}
			\centering
			\includegraphics[width=0.725in]{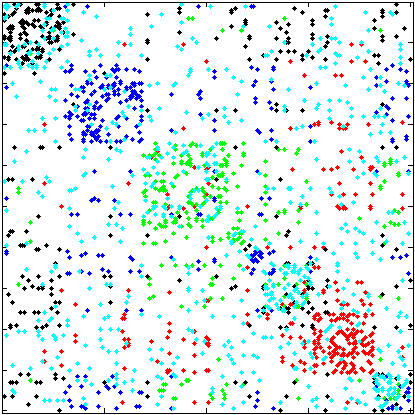}
	\end{minipage}}
	\subfigure[$t = 5$]{
		\begin{minipage}[b]{0.105\textwidth}
			\centering
			\includegraphics[width=0.725in]{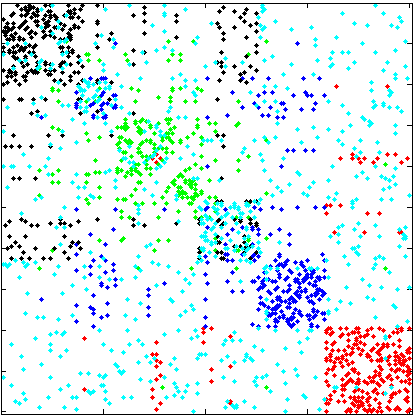}
	\end{minipage}}
	\subfigure[$t = 10$]{
		\begin{minipage}[b]{0.105\textwidth}
			\centering
			\includegraphics[width=0.725in]{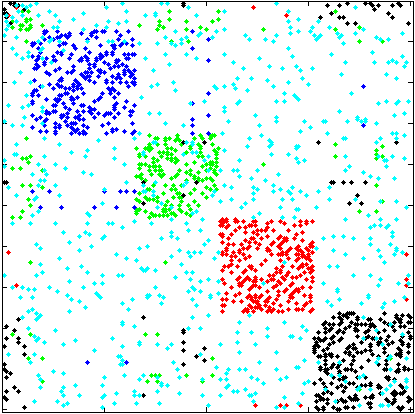}
	\end{minipage}}
	\subfigure[$t = 20$]{
		\begin{minipage}[b]{0.105\textwidth}
			\centering
			\includegraphics[width=0.725in]{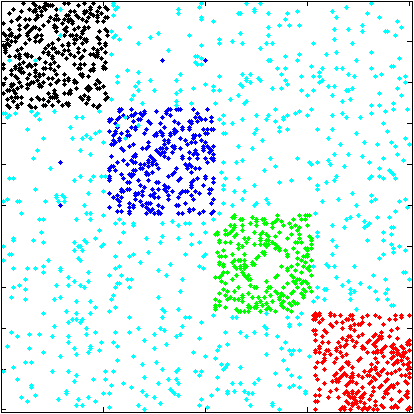}
	\end{minipage}}\vspace{-1.5em}
	\caption{Refinement of layer 2 on SynL2.}
	\label{fig:SynL2_200_layer2} %% label for entire figure
	\vspace{-1.5em}
\end{figure}

There are two planted layers on SynL2, with hiddenness values of $0.45$ and $0.56$.
Communities in layer 2 is comparatively more hidden with respect to communities in layer 1.
%Each of the 5 communities (average size: 40) in layer 1  overlaps with each of the 4 communities (average size: 50) in layer 2 by approximately 10 nodes. 
On average, $20\%$ of the nodes in a community of layer 1 overlap with each community in layer 2, and $25\%$ of the nodes in a community of layer 2 overlap with each community in layer 1.
Figures \ref{fig:SynL2_200_layer1} and \ref{fig:SynL2_200_layer2} illustrate several snapshots of the detected two layers
during the execution (shown by the adjacency matrix but the node IDs are reordered for the two layers respectively).
In the Identification stage ($t=0$), \emph{HICODE} only roughly identifies the stronger layer 1, and only roughly
identifies the less stronger layer 2 after the initial detected layer 1 is weakened.
Then during the Refinement stage, by iteratively weakening the other community layers, we get a more accurate current layer,
which forms a positive feedback cycle. The refinement process converges within 20 steps
and the two detected layers remain stable in further iterations.

\begin{figure*}[htbp]
	\vspace{-1em}
	\subfigure[Caltech]{
		\label{fig:avg_mod_in_RedG_OriG_Caltech} %% label for 1 subfigure
		\begin{minipage}[!]{0.24\textwidth}
			\includegraphics[width=1.4in]{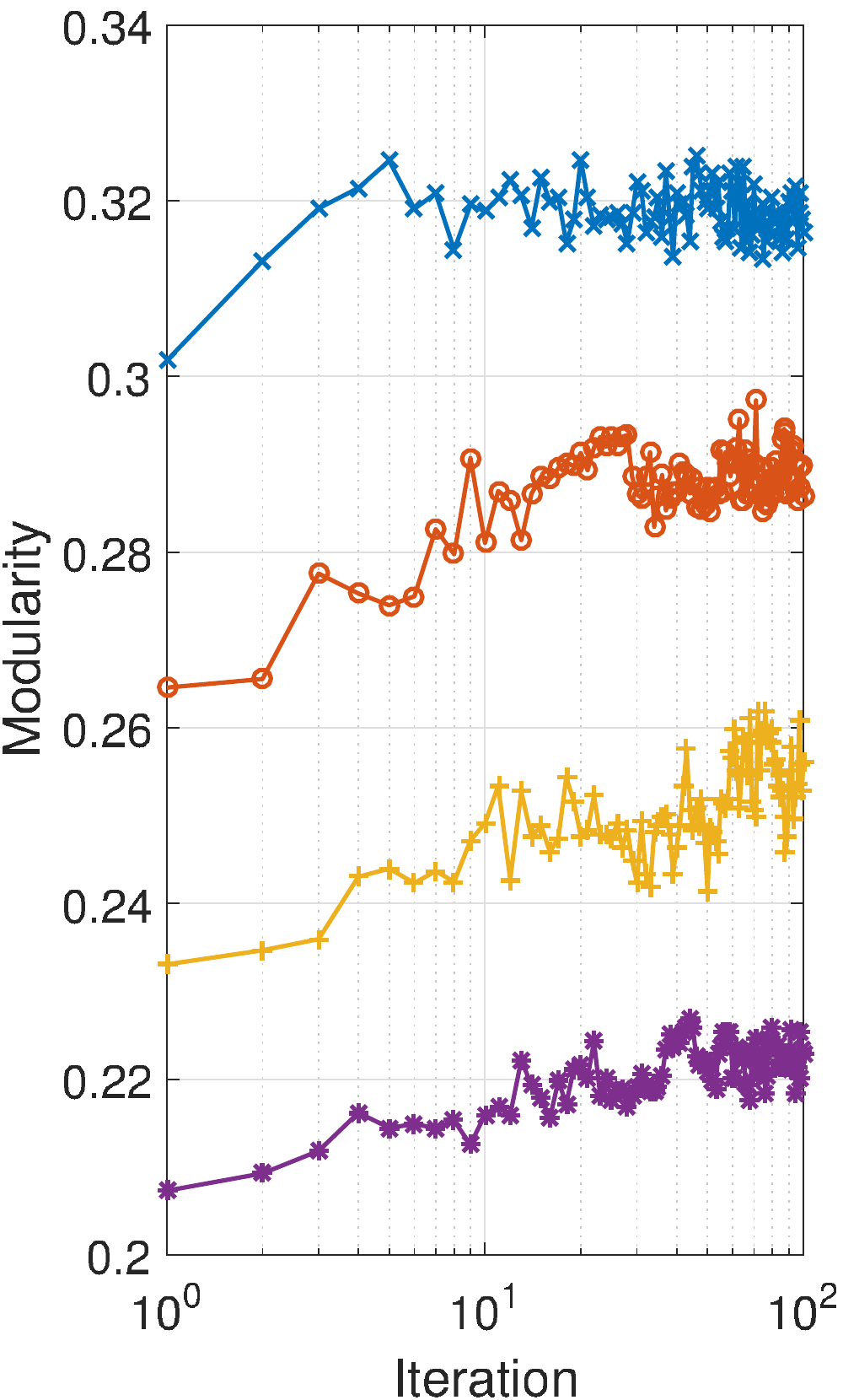} %3.159
	\end{minipage}}
	\subfigure[Vassar]{
		\label{fig:avg_mod_in_RedG_OriG_Vassar} %% label for 5 subfigure
		\begin{minipage}[!]{0.22\textwidth}
			\includegraphics[width=1.3in]{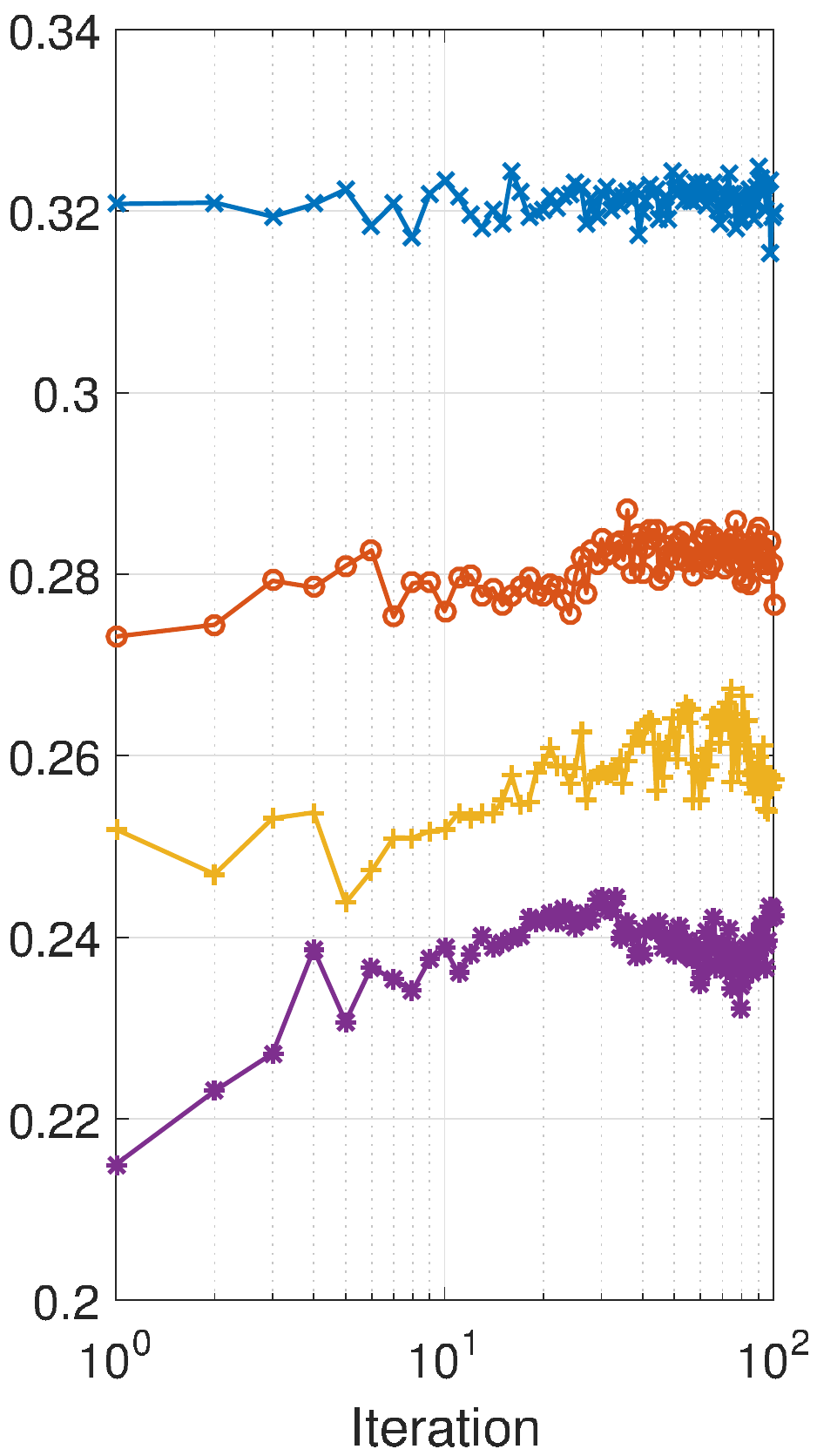}
	\end{minipage}}
	\subfigure[UIllinois]{
		\label{fig:avg_mod_in_RedG_OriG_UIllinois} %% label for 9 subfigure
		\begin{minipage}[p]{0.22\textwidth}
			\includegraphics[width=1.33in]{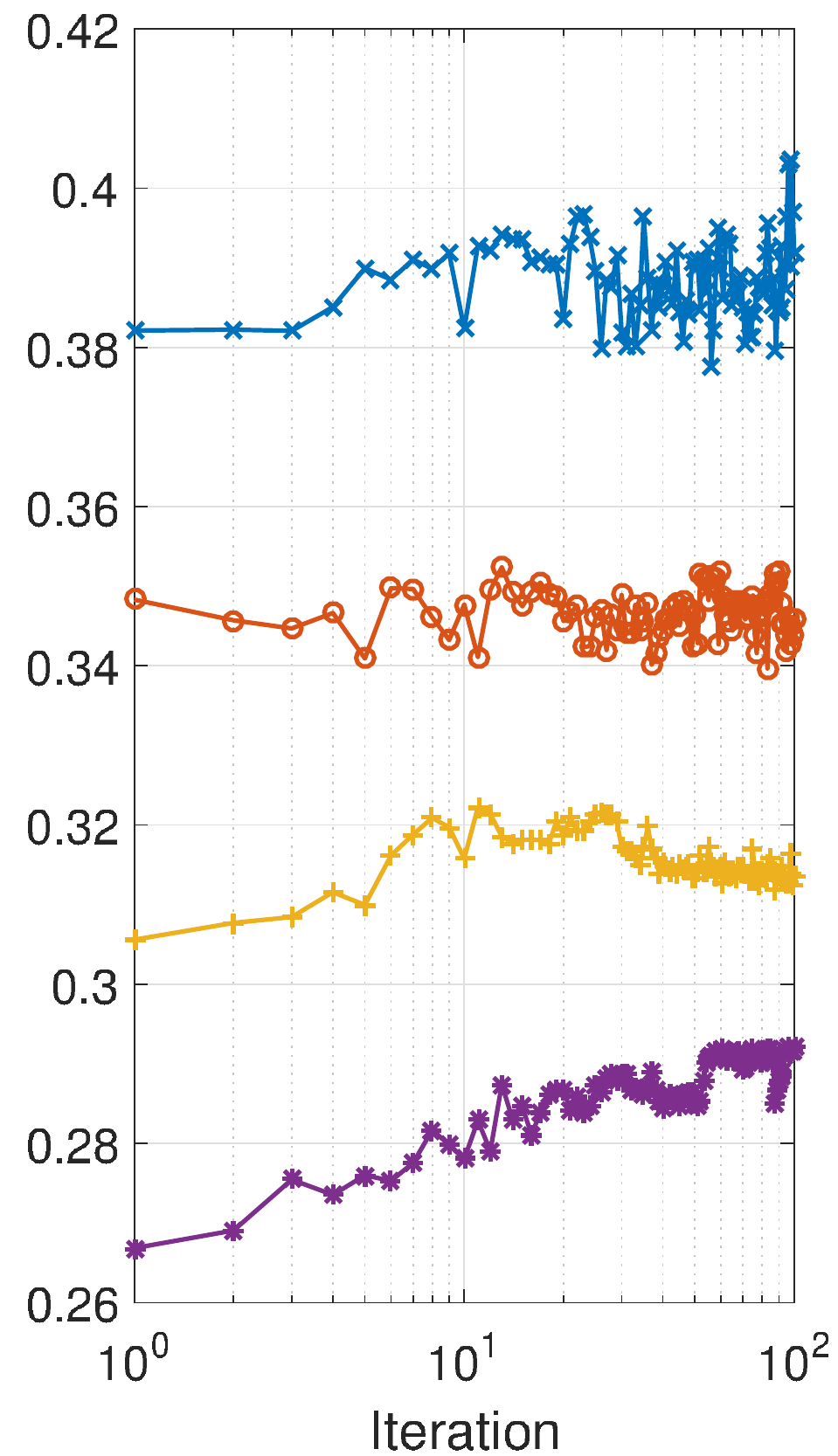}
	\end{minipage}}
	\subfigure[Youtube]{
		\label{fig:avg_mod_in_RedG_OriG_youtube} %% label for 2 subfigure
		\begin{minipage}[!]{0.22\textwidth}
			\includegraphics[width=1.34in]{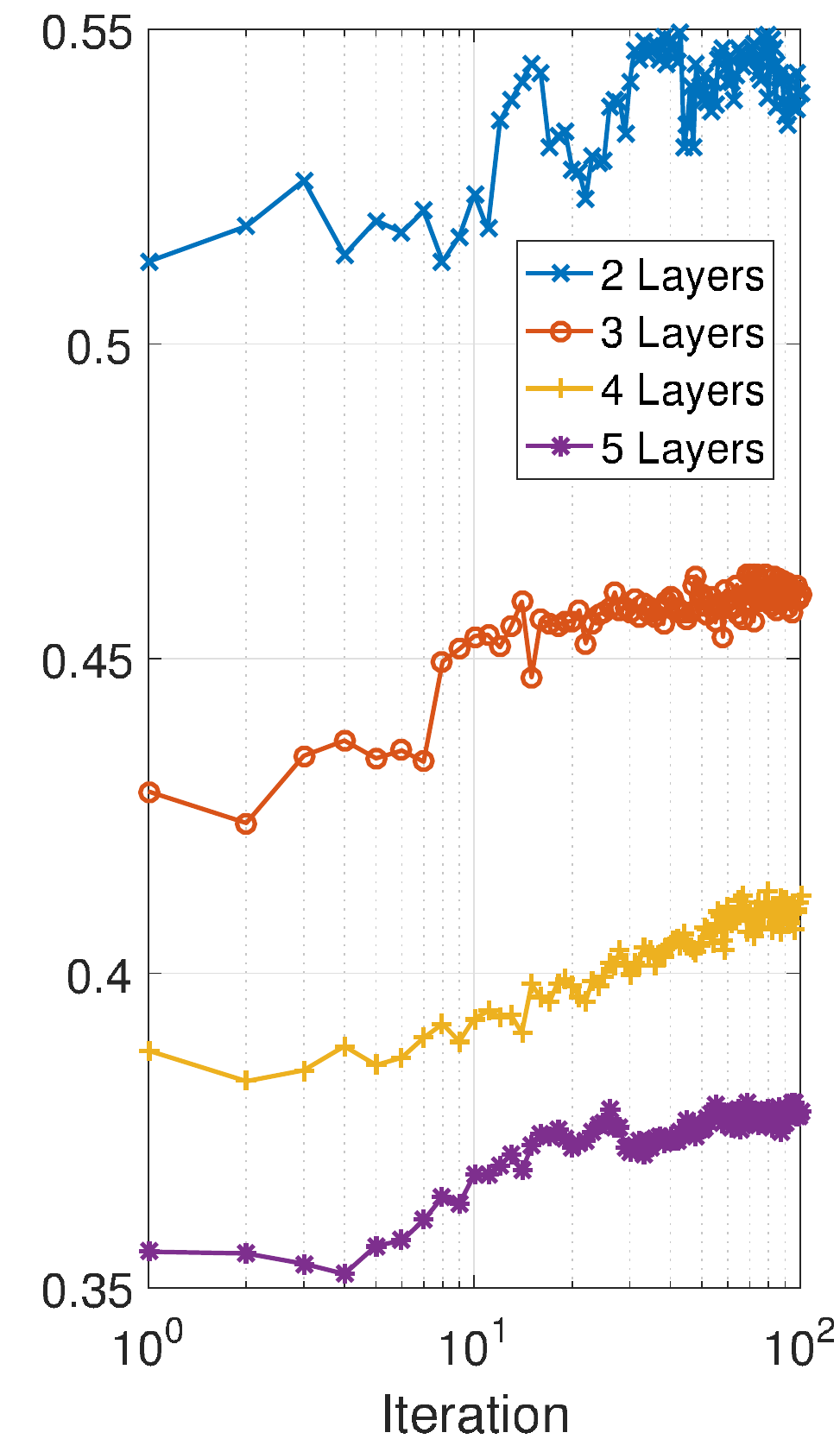}
	\end{minipage}} \vspace{-1em}
	\caption{Improvement on Modularity of the detected layers.} 
	\label{fig:improveMod_on_caltech_youtube} %% label for entire figure
	%\vspace{-1.5em}
\end{figure*}

In addition, we also observe that the average modularity scores of the detected layers on the original network are improved by the Refinement stage. As an example, Figure \ref{fig:improveMod_on_caltech_youtube} shows the increasing trend during the iteration when we detect 2, 3, 4 or 5 community layers on four real-world networks.
%The average modularity of the multiple detected layers increases during the iteration.

%Caltech: 3L, Vassar: 3L, UIllinois: 2L, Youtube: 3L
\setlength{\belowdisplayskip}{-2pt} \setlength{\belowdisplayshortskip}{-2pt}
\setlength{\abovedisplayskip}{-2pt} \setlength{\abovedisplayshortskip}{-2pt}

 % HC:MOD
%\input{060experimentsRealData}

%\vspace{-0.3em}
\subsection{Evaluation on Real Networks}
\label{sec:experiments}
%We apply \emph{HICODE} to both synthetic and real-world networks, as described in Section \ref{sec:ExperimentalSetup}. % as described in Table~\ref{table:network_stats}.

\begin{table*}[htbp]
	\centering
	\small
	\scalebox{0.85}{
		\begin{tabular}{l|  llll| l  l  c  c l}  \hline
			\textbf{Datasets} & \multicolumn{4}{c}{\textbf{Annotated communities}} \vline& \multicolumn{5}{c}{\textbf{Community layers found by {\sc HC:Mod}}} \\
			&Annotations~(modularity, hiddenness value) & $NMI_{max}$ & $F_{max}$ & $F_{avg}$ & \textbf{\#Layers}  & \textbf{Modularity} & $NMI_{max}$ & $F_{max}$ & $F_{avg}$ \\
			\hline
			\textbf{\textit{Facebook}} & & & & & & & & \\
			Caltech  			& Dorm(0.30, 0.08) Year(0.19, 0.84) Status(0.08, 0.85)& 0.13 & 0.32 & 0.18 & 2 & 0.40, 0.25 		 		& 0.00 		& 0.18 & 0.18 \\
			% ONMI(dorm, year)=0.01 ONMI(status, year)=0.01
			Smith 			& Dorm(0.23, 0.42) Year(0.23, 0.49)& 0.00 & 0.04 & 0.04 & 2 & 0.51, 0.34		 		& 0.03		& 0.11 & 0.11\\
			Rice				& Dorm(0.37, 0.02) Year(0.23, 0.94) Status(0.13, 0.91)& 0.11 & 0.26 & 0.14 & 3 & 0.50, 0.36, 0.34   		& 0.02	 	& 0.20 & 0.13\\
			% ONMI(dorm, status)=0.002 ONMI(year, status)=0.11
			Vassar    		& Year(0.34, 0.06) Dorm(0.15, 0.98) Status(0.13, 0.89)& 0.18 & 0.30 & 0.18 & 3 & 0.45, 0.32, 0.31	 		& 0.04		& 0.21 & 0.15 \\
			% ONMI(year, status)=0.17 ONMI(dorm, status)=0.022
			Bucknell			& Year(0.40, 0.05) Status(0.12, 0.91) Dorm(0.11, 0.98)& 0.15 & 0.31 & 0.17 & 3 & 0.51, 0.37, 0.31	 		& 0.05		& 0.30 & 0.22\\
			% ONMI(dorm, status)=0.02 ONMI(year, status)=0.15
			Carnegie			& Year(0.28, 0.29) Major(0.12, 0.84) Status(0.11, 0.78) Dorm(0.08, 0.92)& 0.07 & 0.24 & 0.09 & 4 & 0.40, 0.42, 0.34, 0.32	& 0.06		& 0.23 & 0.17\\
			% ONMI(status, dorm)=0.012 ONMI(status, year)=0.072 ONMI(status, major)=0.027 ONMI(dorm, major)=0.02
			Wellesley			& Year(0.30, 0.10) Status(0.15, 0.79) & 0.15 & 0.28 & 0.28 & 2 & 0.37, 0.26				& 0.00 		& 0.15 & 0.15\\
			UIllinois    		& Year(0.27, 0.24) High\_school(0.15, 0.82) Dorm(0.14, 0.76)& 0.00 & 0.07 & 0.03 & 2 & 0.45, 0.34	 		& 0.04		& 0.16 & 0.14\\
			% ONMI(highschool, dorm)=0.002
			\hline
			\textbf{\textit{SNAP}} & & & & & & & &  & \\
			YouTube  			&  Co-liked $CommSet_A$(0.23, 0.24) $CommSet_B$(0.10, 0.73) &0.03 &0.15 &0.15 & 3 & 0.59, 0.48, 0.33 		& 0.04		& 0.33 & 0.23\\
			Amazon 			&  Co-purchased $CommSet_A$(0.99, 0.29) $CommSet_B$(0.47, 0.92) &0.69 &0.72 &0.72 &  2 & 0.99, 0.14	    	& 0.55		& 0.71 & 0.70\\
			DBLP 				& Co-authorship $CommSet_A$(0.79, 0.13) $CommSet_B$(0.03, 0.59) &0.53 &0.10 &0.10 & 3 & 0.90, 0.72, 0.56		& 0.01		& 0.10 & 0.09\\ \hline
	\end{tabular}} \vspace{0.4em}
	\caption{Statistics of the detected layers found by {\sc HC:Mod} as compared with the annotated communities.}
	\vspace{-1.5em}
	\label{table:all_network_layers}
\end{table*}
% We do not calculate hiddenness values for the networks with only one layer of ground truth communities.

For real-world networks from various domains, 
\emph{HICODE} uncovers multiple layers of high modularity community structure. Table~\ref{table:all_network_layers} (right column) shows the statistics of the layers found by {\sc HC:Mod} in each network. When comparing these layers to one another, their maximum pairwise NMI, maximum and average pairwise $F_1$ scores are low, indicating that these layers of communities are distinct.
One exception is Amazon, which seems only to contain one layer of very strong community structure.

When comparing the ground truth layers on the Facebook networks, 
we see the layers found by \emph{HICODE} are strongly associated with the community categories.
Tables~\ref{table:allF1} show the $F_1$ and $NMI$ scores of \emph{HICODE} and the baseline algorithms evaluated against different annotation categories on several examples: Caltech, Smith, Rice and Vassar.
We see that:

(1)  All baseline algorithms primarily locate the dominant structure, and rarely detect the hidden, weaker structures. Interestingly, different to other algorithms, LC regards Status and Year as the dominant on Vassar and Smith respectively. This occurs as LC may have a different notion of community structure.

(2)  All \emph{HICODE} implementations return community layers that are strongly associated with each ground truth category. \emph{HICODE} not only uncovers the hidden layers that the baseline algorithms rarely detect, but also improves the detection accuracy on the dominant layer.

%\begin{itemize}[noitemsep,nolistsep]
%\item  All baseline algorithms primarily locate the dominant structure, and rarely detect the hidden, weaker structures. Interestingly, different to other algorithms, LC regards Status and Year as the dominant on Vassar and Smith respectively. This occurs as LC may have a different notion of community structure.
%\item  All \emph{HICODE} implementations return community layers that are strongly associated with each ground truth category. \emph{HICODE} not only uncovers the hidden layers that the baseline algorithms rarely detect, but also improves the detection accuracy on the dominant layer.
%%\item  In general, {\sc HC:Mod} and {\sc HC:IM} outperform {\sc HC:OS}, and {\sc HC:OS} outperforms {\sc HC:LC}.
%\end{itemize}

%For the SNAP networks, we partition the ground-truth communities into several community sets ($CommSet$) by their hiddenness value. 
For each $CommSet$ of the SNAP networks, 
we calculate the Recall of \textit{AllLayers} (the union of all layers found by \emph{HICODE}) and the Recall of the communities detected by each of the baselines.
In Table~\ref{table:SNAP},
we see that the comparatively hidden $CommSet_B$ is consistently harder to be located than the dominant $CommSet_A$. 
When compared with the baselines, \emph{HICODE} not only has higher detection accuracy on the comparatively hidden $CommSet_B$,
but also boost the detection accuracy on the comparatively dominant communities in $CommSet_A$.
Again, on Amazon, all algorithms detect both $CommSet_A$ and $CommSet_B$
fairly well due to the highly similarity of the two sets of ground truth communities. 
% because the ground-truth communities are highly overlapped according to 
%Table~\ref{table:all_network_layers},
%which means if only the algorithm found most of the communities in $CommSet_A$ ,
%the detected communities will also have good $NMI$ and $F_1$ score with $CommSet_B$. 
%These experiments demonstrate that \emph{HICODE} is able to detect real, meaningful hidden ground truth layers, while most other algorithms mainly identify the dominant layer.
%%%%%%%%%%%%%%%%%%%%%%%%%%%
\vspace{-1em}
\begin{table*}[htbp]
	\centering
	\small
	\subtable{
		\scalebox{0.9}{
			\begin{tabular}{ l r | c c | c c c |c c | c c | c c | c c}	\hline
				& & \multicolumn{2}{c}{\textbf{HC:MOD}}  \vline&  \multicolumn{3}{c}{\textbf{HC:IM}} \vline& \multicolumn{2}{c}{\textbf{HC:OS}}  \vline & \multicolumn{2}{c}{\textbf{HC:LC}}  \vline & \multicolumn{2}{c}{Partitioning} \vline& \multicolumn{2}{c}{Overlapping} \\
				\hline
				\hline
				\textbf{Caltech}	& 	  & $L_1$ & $L_2$ & $L_1$ & $L_2$ & $L_3$ & $L_1$ & $L_2$ & $L_1$ & $L_2$ & Mod & IM & OS  & LC \\
				\hline
				Dorm & $F_1$ & {\cellcolor{blue!5}\bf{0.58}} & 0.11 & {\cellcolor{blue!5}\bf{0.65}} & 0.11 & 0.11 & {\cellcolor{blue!5}\bf{0.48}} & 0.11 & {\cellcolor{blue!5}\bf{0.21}} & 0.10 & 0.51 & 0.51 & 0.49  & 0.18 \\
				HV = 0.08 & $NMI$ & {\cellcolor{blue!5}\bf{0.39}} & 0.00 & {\cellcolor{blue!5}\bf{0.48}} & 0.01 & 0.01 & {\cellcolor{blue!5}\bf{0.32}} & 0.01 & {\cellcolor{blue!5}\bf{0.16}} & 0.02 & 0.36 & 0.42 & 0.28  & 0.14 \\
				\hline
				Year & $F_1$ & 0.11 & {\cellcolor{blue!5}\bf{0.60}} & 0.12 & {\cellcolor{blue!5}\bf{0.40}} & 0.20 & 0.14 & {\cellcolor{blue!5}\bf{0.45}} & 0.07 & {\cellcolor{blue!5}\bf{0.15}} & 0.13 & 0.14 & 0.12  & 0.07 \\
				HV = 0.84 & $NMI$ & 0.00 & {\cellcolor{blue!5}\bf{0.38}} & 0.03 & {\cellcolor{blue!5}\bf{0.19}} & 0.09 & 0.00 & {\cellcolor{blue!5}\bf{0.29}} & 0.02 & {\cellcolor{blue!5}\bf{0.10}} & 0.05 & 0.13 & 0.00  & 0.03 \\
				\hline
				Status & $F_1$ & 0.17 & {\cellcolor{blue!5}\bf{0.37}} & 0.12 & 0.38 & {\cellcolor{blue!5}\bf{0.64}} & 0.02 & {\cellcolor{blue!5}\bf{0.36}} & 0.23 & {\cellcolor{blue!5}\bf{0.51}} & 0.16 & 0.16 & 0.12  & 0.03 \\
				HV = 0.85 & $NMI$ & \bf{0.16} & {\cellcolor{blue!5}\bf{0.11}} & 0.15 & 0.14 & {\cellcolor{blue!5}\bf{0.32}} & 0.00 & {\cellcolor{blue!5}\bf{0.22}} & 0.12 & {\cellcolor{blue!5}\bf{0.25}} & 0.06 & 0.30 & 0.19  & 0.13 \\
				\hline
				\hline
				\textbf{Smith}	& 	  & $L_1$ & $L_2$ & $L_1$ & $L_2$ & -- & $L_1$ & $L_2$ & $L_1$ & $L_2$ & Mod & IM & OS  & LC \\ \hline
				Dorm	& $F_1$ & {\cellcolor{blue!5}\bf{0.45}} & 0.04 & {\cellcolor{blue!5}\bf{0.50}} & 0.04 & -- & {\cellcolor{blue!5}\bf{0.40}} & 0.07 & 0.05 & 0.01 & 0.25 & 0.43 & 0.38  & 0.04 \\
				HV = 0.42	& $NMI$ & {\cellcolor{blue!5}\bf{0.26}} & 0.00 & {\cellcolor{blue!5}\bf{0.36}} & 0.00 & -- & {\cellcolor{blue!5}\bf{0.25}} & 0.01 & 0.01 & 0.00 & 0.14 & 0.31 & 0.23  & 0.00 \\
				\hline
				Year	& $F_1$ & 0.12 & {\cellcolor{blue!5}\bf{0.56}} & 0.10 & {\cellcolor{blue!5}\bf{0.35}} & -- & 0.15 & {\cellcolor{blue!5}\bf{0.24}} & 0.01 & {\cellcolor{blue!5}\bf{0.20}} & 0.21 & 0.18 & 0.16  & 0.18 \\
				HV = 0.49	& $NMI$ & 0.00 & {\cellcolor{blue!5}\bf{0.37}} & 0.03 & {\cellcolor{blue!5}\bf{0.16}} & -- & 0.05 & {\cellcolor{blue!5}\bf{0.11}} & 0.00 & {\cellcolor{blue!5}\bf{0.03}} & 0.06 & 0.06 & 0.06  & 0.00 \\
				\hline
		\end{tabular}}
	}
	\\% \vspace{-0.5em}
	\subtable{
		\centering
		\small
		\scalebox{0.9}{
			\begin{tabular}{ l r | c c c | c c c | c c c | c c c | c c | c c}   \hline
				& & \multicolumn{3}{c}{\textbf{HC:MOD}} \vline& \multicolumn{3}{c}{\textbf{HC:IM}}  \vline& \multicolumn{3}{c}{\textbf{HC:OS}}  \vline & \multicolumn{3}{c}{\textbf{HC:LC}}  \vline & \multicolumn{2}{c}{Partitioning} \vline& \multicolumn{2}{c}{Overlapping} \\ % \cline{3-15}
				\hline
				\hline
				\textbf{Vassar}    & & $L_1$ & $L_2$ & $L_3$ & $L_1$ & $L_2$ & $L_3$ & $L_1$ & $L_2$ & $L_3$ & $L_1$ & $L_2$ & -- & Mod & IM & OS  & LC \\ \hline
				Year	& $F_1$ & {\cellcolor{blue!5}\bf{0.67}} & 0.16 & 0.10 & {\cellcolor{blue!5}\bf{0.54}} & 0.17 & 0.15 & 0.14 & {\cellcolor{blue!5}\bf{0.44}} & 0.22 & {\cellcolor{blue!5}\bf{0.17}} & 0.08 & -- &
				0.68 & 0.47 & 0.37  & 0.16 \\
				HV = 0.06	& $NMI$ & {\cellcolor{blue!5}\bf{0.47}} & 0.01 & 0.00 & {\cellcolor{blue!5}\bf{0.31}} & 0.07 & 0.02 & 0.06 & {\cellcolor{blue!5}\bf{0.19}} & 0.06 & {\cellcolor{blue!5}\bf{0.04}} & 0.02 & -- &
				0.38 & 0.27 & 0.14  & 0.04 \\
				\hline
				Dorm	& $F_1$ & 0.13 & {\cellcolor{blue!5}\bf{0.41}} & 0.08 & 0.11 & 0.09 & {\cellcolor{blue!5}\bf{0.25}} & 0.11 & {\cellcolor{blue!5}\bf{0.15}} & 0.11 & 0.07 & {\cellcolor{blue!5}\bf{0.08}} & -- &
				0.12 & 0.12 & 0.16  & 0.08 \\
				HV = 0.98	& $NMI$ & 0.03 & {\cellcolor{blue!5}\bf{0.24}} & 0.02 & 0.02 & 0.01 & {\cellcolor{blue!5}\bf{0.08}} & 0.01 & {\cellcolor{blue!5}\bf{0.07}} & 0.02 & 0.02 & {\cellcolor{blue!5}\bf{0.03}} & -- &
				0.00 & 0.02 & 0.00  & 0.00 \\
				\hline	
				Status	& $F_1$ & {\cellcolor{blue!5}\bf{0.34}} & 0.23 & 0.12 & 0.34 & {\cellcolor{blue!5}\bf{0.57}} & 0.19 & 0.52 & 0.23 & {\cellcolor{blue!5}\bf{0.61}} & {\cellcolor{blue!5}\bf{0.63}} & 0.27 & -- &
				0.33 & 0.33 & 0.23  & 0.61 \\
				HV = 0.89	& $NMI$ & {\cellcolor{blue!5}\bf{0.15}} & 0.06 & 0.01 & 0.10 & {\cellcolor{blue!5}\bf{0.21}} & 0.06 & 0.20 & 0.09 & {\cellcolor{blue!5}\bf{0.27}} &  {\cellcolor{blue!5}\bfseries{0.04}} & \bf{0.08}& -- &
				0.15 & 0.10 & 0.07  & 0.04 \\
				\hline
				\hline
				
				\textbf{Rice}    & & $L_1$ & $L_2$ & $L_3$ & $L_1$ & $L_2$ & $L_3$ & $L_1$ & $L_2$ & $L_3$ & $L_1$ & $L_2$ & $L_3$ & Mod & IM & OS  & LC \\ \hline
				Dorm	& $F_1$ & {\cellcolor{blue!5}\bf{0.79}} & 0.11 & 0.07 & {\cellcolor{blue!5}\bf{0.74}} & 0.08 & 0.08 & {\cellcolor{blue!5}\bf{0.61}} & 0.22 & 0.22 & {\cellcolor{blue!5}\bf{0.20}} & 0.10 & 0.07 & 0.70 & 0.55 & 0.54  & 0.10 \\
				HV = 0.02    & $NMI$ & {\cellcolor{blue!5}\bf{0.71}} & 0.00 & 0.00 & {\cellcolor{blue!5}\bf{0.45}} & 0.00 & 0.00 & {\cellcolor{blue!5}\bf{0.50}} & 0.10 & 0.11 & {\cellcolor{blue!5}\bf{0.10}} & 0.01 & 0.00 & 0.59 & 0.32 & 0.29  & 0.04 \\
				\hline
				Year	& $F_1$ & 0.08 & 0.22 & {\cellcolor{blue!5}\bf{0.55}} & 0.08 & 0.20 & {\cellcolor{blue!5}\bf{0.34}} & 0.09 & {\cellcolor{blue!5}\bf{0.22}} & 0.20 & 0.07 & 0.14 & {\cellcolor{blue!5}\bf{0.15}} & 0.07 & 0.08 & 0.14  & 0.05 \\
				HV = 0.94	& $NMI$ & 0.00 & 0.07 & {\cellcolor{blue!5}\bf{0.29}} & 0.00 & 0.06 & {\cellcolor{blue!5}\bf{0.16}} & 0.00 & {\cellcolor{blue!5}\bf{0.13}} & 0.10 & 0.01 & 0.01 & {\cellcolor{blue!5}\bf{0.07}} & 0.05 & 0.05 & 0.00  & 0.00 \\
				\hline
				Status	& $F_1$ & 0.11 & {\cellcolor{blue!5}\bf{0.42}} & 0.23 & 0.10 & {\cellcolor{blue!5}\bf{0.61}} & 0.23 & 0.13 & 0.32 & {\cellcolor{blue!5}\bf{0.58}} & 0.05 & {\cellcolor{blue!5}\bf{0.61}} & 0.12 & 0.10 & 0.08 & 0.14  & 0.03 \\
				HV = 0.91	& $NMI$ & 0.00 & {\cellcolor{blue!5}\bf{0.20}} & 0.11 & 0.01 & {\cellcolor{blue!5}\bf{0.25}} & 0.09 & 0.01 & 0.16 & {\cellcolor{blue!5}\bf{0.42}} & 0.00 & {\cellcolor{blue!5}\bf{0.05}} & 0.02 & 0.04 & 0.01 & 0.00  & 0.01 \\
				\hline
		\end{tabular}}\vspace{-0.1em}
	}
	\caption{Jaccard $F_1$ and $NMI$ scores of all algorithms on Caltech, Smith, Vassar, Rice community categories.}
	\label{table:allF1}
	%\vspace{-1em}
\end{table*}

%\vspace{-0.5em}
\section{Related Work}
\label{sec:relatedWork}
In the past decades, a plethora of community detection algorithms have been presented
for uncovering such latent modular structure based on different metrics (\emph{e.g., modularity~\cite{newman2004Mod,Newman2006Mod,HeKDD2016}, conductivity~\cite{Ncut2000,AndersenWWW2006,Jaewon2012ground}, etc.}) and techniques (\emph{e.g., random walk~\cite{PageRank2006,Rosvall2008Infomap}, heat kernel~\cite{HeatKernel2014}, spectral clustering~\cite{White2005SDM}, seed set expansion~\cite{Palla2005LCFinder,Whang2013seed,Isabel2014Seed}, distance dynamics~\cite{ShaoKDD2015}, mixed membership stochastic blockmodels~\cite{Airoldi2008MMSB, Karrer2011Blockmodel,AbbeNIPS2015}, etc.}). 
Still, there are many new technologies and directions emerging, such as the local community detection~\cite{HeatKernel2014, Isabel2014Seed, He2015ICDM}.  

To the best of our knowledge, this is the first work to formally propose and address the hidden community detection problem. 
Due to space limit, here we only highlight several related works in the area of clustering and community detection. 
For comprehensive reviews to various community detection algorithms and techniques, please refer to survey papers such as~\cite{Coscia2011survey,Xie2013survey}.

Outside the realm of community detection, researchers have studied the problem of clustering data into multiple alternative groupings.
%For example, Niu et al.~\cite{Niu2010Nonredundant, Jordan2014} and Cui et al.~\cite{Cui2007Multiview} use techniques based on spectral and orthogonal clustering to find multiple non-redundant clusterings of the data, where data points of one cluster can belong to different clusters in other views.  
%For high-dimensional data, different feature subspaces may reveal different structures of the data.
By adopting orthogonal clustering and clustering in orthogonal subspaces, Cui et al.~\cite{Cui2007Multiview} cluster the data points in different views, where data points of one cluster can belong to different clusters in other views.
By augmenting a spectral clustering objective function to incorporate dimensionality reduction and multiple views, and to penalize for redundancy between the views,
Niu et al. proposed approaches to learn non-redundant subspaces that provide multiple views simultaneously~\cite{Niu2010Nonredundant} and iteratively~\cite{Jordan2014}.
%with a clustering in each view.
%Their work shows there exist multiple sets of clusterings in different feature subspaces for the high dimension data and is similar in spirit to our work. How to apply our iterative reducing method to strengthen each set of clustering for the high dimensional data, or how to apply their multi-view spectral clusterings for the community detection would be  interesting.  	
In contrast to our work, note that their work is not about finding \textit{hidden} structure, but rather high-quality orthogonal clusterings (possibly in a projected space), which may or may not be hidden by stronger clusterings.  
%Cui:
%Nui: by augmenting a spectral clustering objective function to incorporate
%dimensionality reduction and multiple views and to penalize for redundancy
%between the views. they simultaneously learn non-redundant subspaces that provide multiple views with a clustering in each view

In the realm of community detection, there are several pioneer and embryonic works. Yang and Leskovec~\cite{Yang2012ICDMb} found that community overlaps are more densely connected than the non-overlapping parts, which coincide with our assumption that the overlapping part are the superposition of the layers. And they propose a Community-Affiliation Graph model to maximize the affiliation likelihood.
Then, there are three pieces of related work that are proposed independently and almost simultaneously~\cite{Chen2014Deep, Young2015Hidden, He2015arXiv}.

Chen et al.~\cite{Chen2014Deep} remove nodes or edges based on the local Fiedler vector centrality (LFVC) which is associated with the sensitivity of algebraic connectivity to node or edge removals. Most importantly, they define a concept of deep community as a connected component that can only be seen after removal of all nodes or edges from the rest of the network. They prove that their method works on small synthetic networks under stochastic block model framework and do experiments on small networks of size hundreds.

Young's work~\cite{Young2015Hidden} is most similar in spirit to our work, and they also reference our first version of the work~\cite{He2015arXiv}. They observe that smaller or sparser communities can be `overshadowed' by the larger or denser communities, and communities may appear at different resolutions. 
They use two existing algorithms as base algorithms (LC~\cite{Ahn2010LinkCommunities} and CFinder~\cite{Palla2005LCFinder}),  find the first set of communities, remove all internal edges for the current detected communities, find a second set and repeated the process until no significant communities could be found. Note that communities are fixed once extracted, and they did not really uncover the weak, hidden communities that actually are incoherent with the dominant communities. Their method corresponds to the Identification stage of {\sc HC:LC} using RemoveEdge.
The detailed comparison is in the early version of our paper uploaded to arXiv~\cite{He2015arXiv}.

\begin{table}[htbp]
	%\vspace{-1em}
	\centering
	\small
	\subtable{
		\scalebox{0.78}{
			\begin{tabular}{ l r | c | c | c | c | c c | c c}	\hline
				\textbf{Youtube}	&	& \textbf{HC:MOD} & \textbf{HC:IM} & \textbf{HC:OS} & \textbf{HC:LC} & 
				Mod & IM & OS  & LC \\
				\hline
				$CommSet_A$ &	& \bf{0.16} & \bf{0.27} & \bf{0.30} & \bf{0.28} & 
				0.13 & 0.04 & 0.27 & 0.28 \\
				\hline
				$CommSet_B$ &	& \bf{0.10} & \bf{0.12} & \bf{0.13} & \bf{0.13} & 
				0.05 & 0.03 & 0.08 & 0.12\\
				\hline
				\hline
				\textbf{Amazon}	&	& \textbf{HC:MOD} & \textbf{HC:IM} & \textbf{HC:OS} & \textbf{HC:LC} & 
				Mod & IM & OS  & LC \\
				\hline
				$CommSet_A$ &	& \bf{0.93} & \bf{0.90} & \bf{0.90} & \bf{0.91} & 
				0.89 & 0.88 & 0.78 & 0.83 \\
				\hline
				$CommSet_B$ &	& \bf{0.90} & \bf{0.86} & \bf{0.89} & \bf{0.88} & 
				0.86 & 0.83 & 0.82 & 0.79\\
				\hline
				\hline
				\textbf{DBLP}	&	& \textbf{HC:MOD} & \textbf{HC:IM} & \textbf{HC:OS} & \textbf{HC:LC} & 
				Mod & IM & OS  & LC \\
				\hline
				$CommSet_A$ &	& \bf{0.23} & \bf{0.21} & \bf{0.29} & \bf{0.22} & 
				0.19 & 0.19 & 0.21 & 0.15\\
				\hline
				$CommSet_B$ &	& \bf{0.11} & \bf{0.15} & \bf{0.20} & \bf{0.24} & 
				0.09 & 0.09 & 0.14 & 0.18\\
				\hline
		\end{tabular}}
	}
	\caption{Jaccard Recall $R$ of all algorithms on SNAP data.}
	\label{table:SNAP}
	%\vspace{-2em}
\end{table}

% \vspace{-2em}

\section{Conclusions}

In this paper, we propose a new paradigm of hidden community structure for network analysis and communuty detection. 
We present a formal definition on `hidden structure', and propose a systematic method called \textit{HICODE} to uncover these hidden communities.
Through extensive experiments, we demonstrate that hidden communities exist in real-world networks of various domains, and the proposed method significantly outperforms several comparative methods, including overlapping community detection algorithms, in accurately locating these hidden structure.
Our work sheds light on the organization of complex networks and provides new directions for research on community detection.

In the future, we plan to explore the hidden communities on directed or weighted networks, apply other community detection algorithm as the base, as well as design new reduction method for weakening the detected communities. 
It is also possible that our new ideas can be applied to other network mining tasks such as link-prediction or maxmimum influence propagation. 
 
%\vspace{-1em}
\section*{Acknowledgments}
%The work is supported by US Army Research Office (W911NF-14-1-0477), NSFC (61472147), Microsoft Research Asia Collaborative Research (97354136), and NSFC of Hubei Province (2015CFB566).
The work is supported by US Army Research Office (No. W911NF-14-1-0477), National Science Foundation of China (No. 61472147), Microsoft Research Asia Collaborative Research (No. 97354136), and National Science Foundation of Hubei Province (No. 2015CFB566).

%\appendix
%Appendix A
%\section{Related Metrics}
%\label{sec:preliminaries}
%\input{030preliminaries}

%\bibliographystyle{ACM-Reference-Format}
\bibliography{HiCodeKDD}

\end{document}